\documentclass[aps,prb,twocolumn,floatfix,nobibnotes,groupedaddress,amsmath,amssymb,amsfonts]{revtex4}

\usepackage{graphicx}
\usepackage{latexsym}

\usepackage{float}
\begin{document}

\title{A versatile maskless
 microscope projection photolithography system and its application in light-directed fabrication of DNA microarrays}

\author{Thomas Naiser}
\email{thomas.naiser@ep1.uni-bayreuth.de}
\affiliation{Experimentalphysik I, Universit\"at Bayreuth, D-95440 Bayreuth, Germany}

\author{Timo Mai}
\affiliation{Experimentalphysik I, Universit\"at Bayreuth, D-95440 Bayreuth, Germany}

\author{Wolfgang Michel}
\affiliation{Experimentalphysik I, Universit\"at Bayreuth, D-95440 Bayreuth, Germany}

\author{Albrecht Ott}
\affiliation{Experimentalphysik I, Universit\"at Bayreuth, D-95440 Bayreuth, Germany}

\date{\today}

\begin{abstract}

The manuscript has been published in Review of Scientific Instruments (http://rsi.aip.org/rsi/) Rev. Sci. Instrum. 77, 063711 (2006)
\newline
\newline
We present a  maskless microscope projection lithography system (MPLS), in which  photomasks have been replaced by a Digital Micromirror Device type spatial light modulator (DMD\textsuperscript{TM}, Texas Instruments). Employing video projector technology high resolution patterns, designed as bitmap images on the computer, are displayed using a micromirror array consisting of about 786000 tiny individually addressable tilting mirrors. The DMD, which is located in the image plane of an infinity corrected microscope, is projected onto a substrate placed in the focal plane of the microscope objective. With a 5$\times$ (0.25 NA) Fluar microscope objective, a fivefold reduction of the image to a total size of 9 mm\textsuperscript{2} and a minimum feature size of 3.5 $\mu$m is achieved. The ultra high pressure (UHP) lamp of a video projector is a  cheap, durable and powerful alternative to the mercury arc lamps commonly used in lithography applications. The MPLS may be employed in standard photolithography, we have successfully  produced patterns in 40 $\mu$m films of SU-8 photoresist, with an aspect ratio of about 1:10. Our system can be used in the visible range as well as in the near UV (with a light intensity of up to 76 mW/cm\textsuperscript{2} around the 365 nm Hg-line). We developed an inexpensive and simple method to enable exact focusing and controlling of the image quality of the projected patterns. Our MPLS has originally been designed for the light-directed in situ synthesis of DNA microarrays. One requirement is a high UV intensity to keep the fabrication process reasonably short. Another demand is a sufficient contrast ratio over small distances (of about 5 $\mu$m). This is necessary to achieve a high density of features (i.e. separated sites on the substrate at which different DNA sequences are synthesized in parallel fashion) while at the same time the number of stray light induced DNA sequence errors is kept reasonably small. We demonstrate the performance of the apparatus in light-directed DNA chip synthesis and discuss its advantages and limitations. 
\end{abstract}


\maketitle

\section{Introduction}
Since the invention of the integrated circuit, photolithography is one of the most important technologies in semiconductor industry. For photoresist patterning usually chromium masks are used as reticles. The high cost of chrome masks and long turnaround times from mask design to fabrication of the masks are often limiting the application of photolithographic techniques.
In  many scientific fields, especially biosciences, there is a growing interest in using lithographic methods.  Photolithographic techniques have already been applied for the fabrication of biosensors, microfluidic systems ("lab on a chip"), electrode structures, patterning of cell culture substrates\cite{Luebke2004} and for the light-directed fabrication of DNA microarrays,\cite{Fodor1991,Singh-Gasson1999,Luebke2002} just to name a few.
An extensive review on soft lithography applications in biology and biochemistry is given by Whitesides \textit{et al.}.\cite{Whitesides2001} 

As a more affordable alternative to chromium masks, transparent films with patterns printed on a high resolution printer can be used in many applications.\cite{Duffy1998} Employing microscope projection photolithography\cite{Brady83,Love01,Palmer1973,Vandenberg1978,Smith86} to project these patterns onto a photoresist coated surface, microstructures with features as small as 1 $\mu$m have been produced.\cite{Love01}    

In this paper we describe the design and performance of a microscope projection photolithography system (MPLS) employing a DMD spatial light modulator (SLM) for pattern generation. MPLS was designed to meet the requirements of the light-directed synthesis process for microarrays,\cite{Fodor1991} but the versatile system can just as well be applied in other photolithography applications. A microarray, as described in more detail in Sec.~\ref{sec:LightDirectedSynthesis}, comprises a large number of domains (in the context of DNA chips also called features), each of which contains identical single stranded DNA molecules of a different sequence. Typically the number of  features is in the range of 10\textsuperscript{2} to 10\textsuperscript{5} for spotted microarrays (where droplets containing prefabricated DNA are deposited onto the substrate surface), but more than a million features (as on the Affymetrix GeneChip\textsuperscript{TM}) can currently be fabricated on a DNA chip by using photolithographic techniques. Usually glass is used as a support material. The surface mounted probes  act as scavengers  for complementary target DNA molecules, which are applied to the microarray surface in solution. Through base pairing the probes act as highly sensitive detectors for fluorescently labeled nucleic acids containing the complementary sequence. Microarrays with feature sizes ranging from 10 $\mu$m to several hundred microns are increasingly employed in research and diagnostic applications, mainly for the analysis of gene expression.

Singh-Gasson \textit{et al.}\cite{Singh-Gasson1999} were first to apply micromirror technology for replacement of the physical chrome masks. Similar approaches were reported by Luebke \textit{et al.}\cite{Luebke2002}, Nuwaysir \textit{et al.}\cite{Nuwaysir2002} and  Kim \textit{et al.}.\cite{Kim2004} Gao \textit{et al.}\cite{Gao2001} employed DLP technology for  DNA microarray synthesis on microstructured substrates, with a different photochemistry using photogenerated acids to remove acid-labile protection groups. 
Our approach to microarray fabrication focuses on a size-reduction of the microarrays (with an increased feature density). Due to the smaller dimensions of the microarray not only the amount of expensive biochemical ingredients  for the in situ synthesis is reduced, but also the performance  of the diffusion-driven hybridization process is expected to improve.\cite{Wei2005} 
In light-directed synthesis  of microarrays the use of the DMD  not only avoids the expense for the chrome masks but also circumvents the technical problems related to mask alignment. Using this flexible technology the turnaround time from designing a microarray to its application is reduced to about one day. Thus an iterative approach for the optimization of microarray designs is possible. Compared with liquid crystal SLMs (as also commonly used in video projection systems), the  DMDs  have superior image contrast and, unlike the liquid crystal technology, DMDs may also be used in the near UV range.

The main requirements for the light-directed DNA synthesis process are a sufficient light intensity in the near UV around 365 nm (a dose of 7.5 J/cm\textsuperscript{2} is required for  a complete exposure of the photolabile NPPOC-groups\cite{Nuwaysir2002}) and a sufficient image contrast. A local image contrast of the order of 1:100 or better (as outlined below) is desired. Stray light leaking over short distances from exposed features to not exposed ones nearby (e.g. due to astigmatism, spherical aberration, or diffraction at the edges of the micromirrors\cite{Kim2004}) can introduce base insertion errors and can therefore affect the performance of the microarray.
Furthermore, thermal and mechanical stability of the optical system are an important issue. Depending on the length of the oligonucleotide sequences to be synthesized, up to 100 exposure steps (four exposure steps per nucleotide) are necessary. Good alignment of the mask patterns is required. To succesfully synthesize features at the highest (pixel) resolution achievable with our setup (3.5 $\mu$m) the maximum alignment error should be no larger than one  micron.


\section{System Configuration}
The basic idea of MPLS is to use a spatial light modulator for displaying computer generated "virtual photomasks" in the focal plane of a microscope. For that purpose we make use of a DMD spatial light modulator and  its driver electronics, both obtained from a secondhand commercial video projector (Anders+Kern AstroBeam 540). The optics of a Zeiss Axiovert 135 inverted microscope is employed in a reverse optical path for image projection: Using a 5$\times$ (0.25 NA) Fluar microscope objective (Zeiss), the image of the DMD - which is located in the intermediate image plane - is scaled down to  3.5 mm$\times$2.6 mm. As a light source for visible and near UV wavelengths we employ a 250 W Ultra High Pressure (UHP) mercury arc lamp also from a video projector (Optoma EP758). The hardware described in the following sections is illustrated in Fig.1. Components shortcuts refer to annotations on Fig. 1(b).

\begin{figure}[htbp]
  \centering
    \includegraphics{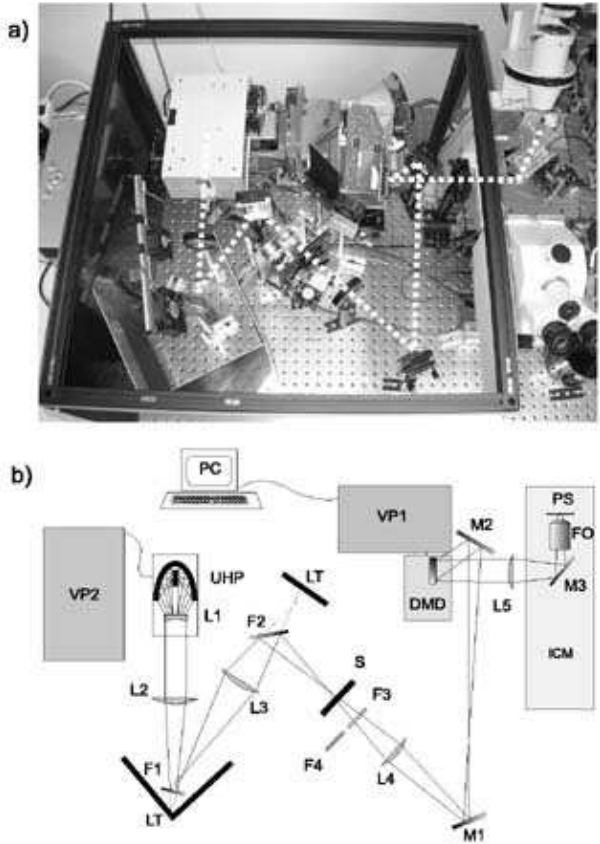}
  \caption{
(a) Photograph of the maskless microscope projection photolithography system (top view). Along the optical path (dotted white line): UHP lamp housing, UV cold mirrors, shutter, band pass filters (green and UV), DMD and driver electronics, tube lens, microscope (Zeiss Axiovert 135) and the reaction cell, which is mounted onto the sample holder. \newline (b) Drawing of the lithography system: Ultra High Pressure lamp (UHP) powered by video projector (VP2), plano-concave silica lens (L1), plano-convex lens (L2), UV cold mirror F1, light trap (LT), plano-convex lens (L3), UV cold mirror (F2), shutter (S), bandpass filters for UV (F3) and green (F4) illumination, plano-convex lens (L4), fold mirrors (M1 and M2), DMD and driver electronics of the Astrobeam projector (VP1), tube lens (L5), infinity corrected microscope (ICM), mirror/beamsplitter-assembly (M3), 5$\times$ (0.25 NA) Fluar microscope objective (FO), substrate to be patterned (PS).
  }
  \label{Labelname}
\end{figure}

\subsection{Illumination system}
Ultra High Pressure  mercury arc lamps are optimized for use in video projection systems: They have an outstanding arc luminance and a very long life of up to 10000 h. The operation pressure of up to 300 bar causes considerable line broadening - resulting in an almost continuous spectrum, which is  advantageous for display applications but can be a problem  in photolithographic applications (for example due to increased chromatic aberrations). 

In our lithography setup the use of a UHP lamp has several advantages over a conventional mercury arc lamp. Compared to the latter, the UHP lamp has a smaller and brighter arc region (with an  arc gap of typically 1-1.5 mm), resulting in an increased light throughput.  The UHP lamp has a higher arc stability and has a very long life time of several thousand hours.

Due to the requirement for high UV transmission we couldn't use the highly optimized optics of the video projector. Therefore the illumination optics had to be redesigned: A replacement lamp module for the Optoma EP758 projector was built into an air cooled housing and connected via an extension cable to the lamp driver of the Optoma projector (VP2). The arc of the 250 W UHP lamp is located at the inner focal point of the elliptical lamp reflector. To efficiently collimate the strongly divergent beam, a plano-concave diffraction lens (L1)(25.4 mm diam., f=50 mm, fused silica) is placed between the lamp window and the outer focal point of the reflector. 

The dichroic mirror of the Optoma lamp module (F1) (originally designed as a UV protection filter) is employed as a UV cold mirror to cut down the visible light intensity to about 10 percent. UV light below 400 nm is efficiently reflected. Infrared radiation is filtered using another UV cold mirror (Oriel) (F2). Finally a band pass interference filter (F3) (bk-370-35-B, Interferenzoptik Elektronik GmbH) is used for selecting the wavelength band in the mercury i-line region ($\lambda=365$ nm) required for the photodeprotection reaction. Taking into account that the mercury i-line is considerably broadened due the high operation pressure of the lamp, we have to use a relatively wide band filter (FWHM: 33 nm) to achieve a sufficiently high UV transmission.

\subsection{Pattern generation using a Digital Micromirror Array}
The DMD is a spatial light modulator commonly used for image generation in DLP video projection systems. In our setup we use a DMD with XGA resolution containing 1024$\times$768=786432 square mirrors (16 $\mu$m in size with a pitch of 17 $\mu$m) that can be tilted by an angle of +$10^{\circ}$ or -10$^{\circ}$ relative to the normal axis of the chip. The two positions are referred to as on- and off-state:  Mirrors in the on-state reflect the incident light perpendicular to the DMD surface into the projection optical system, whereas  mirrors in the off-state reflect light at an angle of $40^{\circ}$ relative to the DMD normal axis into a light trap. 

The DMD is oriented perpendicular to the optical axis of the  projection system. The micromirrors tilt around their diagonal axis. We have rotated the DMD by $45^{\circ}$ around the optical axis, so that the incident beam and the reflected beam lie both in the horizontal plane of the setup. 
For better accessibility of the micromirror array  the DMD board had to be removed from the projector chassis and reconnected to the driver board via a 148 pin extension cable. Because the driver  electronics of the projector remains unchanged, all sorts of video signals can be used to control the image display. Connection to a PC with a dual-head graphics card proved to be useful, as one screen can be used for control purposes (e.g. for running the DNA synthesis control program which automates and coordinates photolithographic pattern display and the fluidics system) while the other one is reserved for pattern display.

\subsection{The image projection optics}
To reduce the size of our microarrays to a few mm\textsuperscript{2} we opted for a microscope projection approach. There are several benefits: By reducing the image area, the illumination intensity is increased by a similar factor, thus reducing the need for increased lamp power. A 250 W lamp does suffice in order to keep the time required for optical deprotection in a reasonable relationship to the total turnover time of the chip synthesis. Use of the microscope also provides superior control of the image focusing and mechanical stability. 

An important aspect in the design of the lithography system is image contrast. In light-directed microarray synthesis stray light (the term is used here for all kind of misguided light) is much more critical than for example with photoresist. Photoresist has a stronly nonlinear exposure characteristics and doesn't respond to small stray light intensities below a threshold value. In microarray synthesis, there is no such threshold and stray light induced errors can accumulate over many exposure steps. Within the total exposure time of about two hours, stray light causes base insertion errors, affecting most of the synthesized DNA strands. 

The whole synthesis process, involves about 80 exposures with different mask patterns, it extends over about 6.5 hours. So there is a demand for thermal and mechanical stability. To make use of the maximum pixel resolution of the setup (which is 3.5 $\mu$m) no movements caused by vibrations, tension release, or thermal expansion larger than about 1 $\mu$m (in the front focal plane of the objective) can be tolerated. 

The micromirror array is placed in the image plane (located outside the microscope frame) of the inverted microscope. With  infinity corrected microscope objectives, a tube lens is necessary to project the image of the DMD to infinity. The adjustment of the distance between  DMD and tube lens, which does not exactly equal the nominal focal length of 164.5 mm (as specified by the manufacturer), is very crucial for the calibration of the setup, as explained later (in Sec.~\ref{Sec:ChromCorrection}). 

A movable half mirror/half beamsplitter optical element (M3), located at the position of the microscope's fluorescence filter block, is used to reflect the light into the objective back aperture.  By making use of the beamsplitter, the light reflected back from the surface of the pattern substrate can be coupled into the microscope, therefore allowing exact focusing and direct observation of the image through the eyepiece. Subsequently, for photopatterning the mirror part is used. In principle for this purpose a dichroic beamsplitter could be used, but then image artifacts  due to reflections from the backside of the beamsplitter need to be eliminated.

Among several objectives tested, we found the Zeiss Fluar 5$\times$ (0.25 NA) as most suitable for DNA chip fabrication, particularly for its superior UV transmittance and its large back aperture allowing for efficient light collection. Over a working distance of 12.5 mm the image of the DMD is projected onto the DNA synthesis substrate - a chemically modified glass surface - inside the reaction cell. 
 
A 10$\times$ (0.30 NA) Plan Neofluar and a 20$\times$ (0.5 NA) Plan Neofluar objective (Zeiss) were successfully used to further reduce the image size. Diminished contrast makes these objectives less suitable for light directed microarray fabrication. However, patterning of photoresist - having lower requirements on contrast - should be simple with these higher magnification objectives. 

At a wavelength of 365 nm the diffraction limit of the 5$\times$ (0.25 NA) objective is 0.73 $\mu$m. A significantly larger distance between adjacent features is necessary to achieve a sufficient local contrast for the light-directed fabrication process.

\section{Methods}
\subsection{Fabrication and application of UV-sensitive photochromic films}
\label{sec:SpiroFab}
For evaluation of the imaging quality a fast and simple method for generating patterns upon UV exposure is required. Photographic films and photoresist turned out to be not very useful due to difficult handling and processing efforts. Therefore we have developed a UV-sensitive film based on the photochromic dye spiropyran. Spiropyran undergoes a structural change when exposed to UV-light. This results in a strongly increased light absorption in the visible range. 

Preparation of photochromic films was performed in the following way:
We dissolved 10 mg of spiropyran dye (1',3'-dihydro-1',3',3'-trimethyl-6-nitrospiro[2H-1-benzopyran-2,2'-(2H)-indole], Aldrich, Cat.: 27,361-9) in  1 ml of PMMA photoresist(E-beam resist PMMA 200 k; AR-P 641.04, Allresist GmbH, Strausberg, Germany) and spincoated a thin film (thickness about 1 $\mu$m) onto a microscope slide. Other resists - we also tried with MicroChem PMMA and MicroChem SU-8 50 - work equally well. The photoresist is used as a carrier material only. After spincoating, and brief heating on a hot plate (1 minute at 100$^{\circ}$C) the slides are ready for use.

We found these photochromic films to be a well-suited imaging material.  
Unlike with photoresist or photographic material no developing or other processing is required.
Under UV exposure the film changes from transparent to an almost opaque purple. With the intensities we usually apply (50-100 mW/cm\textsuperscript{2}) this happens within seconds. The process can be reversed by heating or by illumination with bright light (at visible wavelengths). Unless the  spiropyran has been bleached with high irradiation doses, the films can be reused several times.

As a response to UV light the optical density strongly decreases. For a small exposure dose the optical density increases almost linearly with the dose of UV light. For larger doses D the optical density $OD=OD_{sat}(1-\exp{(-const\cdot D))}$ approaches saturation. Upon very high exposure, photodegradation of the photochromic dye results in reduced OD values.

\subsection{Chromatic correction of the projection optical system}
\label{Sec:ChromCorrection}
Since the depth of focus DOF=$\lambda$/NA\textsuperscript{2} is only about 6 $\mu$m for the 5$\times$ (0.25 NA) Fluar objective (at $\lambda=$365 nm), it is necessary to perform proper focusing each time a new patterning substrate is mounted on the sample holder. The focus range providing optimum contrast is even smaller than the depth of focus, thus perfect focusing of the pattern onto the surface is crucial. It can be achieved by observing the back reflection of the projected image (from the patterning surface) through the microscope eyepiece. This is easy to perform with visible light, but almost not feasible with UV light (even with a UV sensitive CCD camera - due to the restricted size of the CCD chip - one can only see a small part of the image). 

If the back-reflected image of the pattern is perfectly focussed in green light, this usually is not true for UV at the same time. This is due to chromatic aberration. Longitudinal chromatic aberration causes an axial focus shift\cite{Brady83} usually resulting in a completely blurred image in UV. 
In the following we describe a method for the correction of this longitudinal chromatic aberration, so that focusing of the near UV image can be performed by observation (through the eyepiece) and focus adjustment under green light illumination.

Using photochromic films as a control for the quality of the projected UV pattern, we found that the chromatic aberrations can be compensated by fine-adjustment of the distance d between the DMD and the tube lens (see Fig. 1). The distance d is roughly the nominal focal length of the tube lens of 164.5 mm. After focusing with green light, the film is exposed with a control pattern in UV and subsequently inspected on a light microscope. The distance d now can be adjusted iteratively  until the patterns imaged on the spiropyran slide indicate perfect focusing. Just a small deviation of a few millimeters from the nominal focal length of the tube lens is necessary for chromatic correction. The tolerance of d, within which a good correction is achieved, is only a few tenth of a millimeter wide. Once the chromatic correction procedure has been accomplished, focusing can always be performed under illumination with green light.


\section{Experimental results and discussion}

\subsection{UV light intensity and uniformity of illumination}
For measuring the intensity at the image plane we used a laser power sensor (PS10Q, Coherent Inc.). The thermopile sensor was placed in the focal plane of the microscope objective. To measure the mean intensity, a completely white image was displayed on the DMD. With the measured total power of 7.8 mW we determined the intensity in the image plane as 87 mW/cm\textsuperscript{2}. 

To study  the uniformity of the illumination we projected the image onto a screen. The intensity was measured at different regions of the projected image. An asymmetric large scale deviation with a peak intensity of about 140 percent of the mean intensity is observed. This is due to the configuration of the illumination system: The UHP lamp's arc gap is oriented parallel to the optical axis, providing a very inhomogeneous illumination profile. For this reason  in a video projection system an integrator element, e.g. an integrator rod (which is a light guide with a rectangular cross section) or a fly-eye lens array is employed to generate a very uniform illumination. Using the integrator rod of the Astrobeam projector turned out to be not feasible as the glass rod absorbs most of the UV light. 

We decided to flatten the illumination profile by using only a small homogeneous section of the light cone for illuminating the DMD. This way we  sacrifice about 80 percent of the light. Nevertheless, the remaining 20 percent of light allow photodeprotection to be performed in a reasonable time. Alternatively, if such parts were available, a quartz integrator rod or an integrator plate (fly-eye lens array) could be used to achieve significantly higher light intensities.

To attain a more uniform illumination we employ the DMD for intensity leveling, similar as described by Huebschman \textit{et al.}.\cite{Huebschman2004} For this purpose we have created an "intensity leveling mask". The black and white images (to be used as a photomasks) can easily be leveled to reduce intensity variations to about $\pm$10 percent by pixelwise multiplication with this mask. To generate the intensity leveling mask, a fully illuminated image (all mirrors in the on-state) is  projected  on the screen (as described above, still without using the microscope objective) and photographed with a Nikon Coolpix 4500 digital camera. Deskewing the raw image using standard image processing software results in a 1024$\times$768 pixel image, which finally has to be inverted and adjusted in brightness and contrast. The leveling mask is then projected onto the screen and a photometer is used to measure uniformity of illumination. In an iterative way image brightness and contrast are adjusted to achieve a uniform intensity within most of the image area. Contour plots of the light intensity before and after intensity leveling are shown in Fig. 2. Only in the outermost corners of the image (comprising about 10 percent of the total image area) the intensity is reduced to about 50 percent of the mean intensity. This is due to vignetting: Light reflected from the corners of the DMD, which are located close to the edge of the entrance pupil, is partially blocked by the apertures of the tube lens respectively the microscope objective. Applying intensity leveling we achieved a mean light intensity of 76 mW/cm\textsuperscript{2}.

The intensity values mentioned above were achieved using an interference filter with a FWHM of 33 nm and a maximum transmission of 60 percent at a center wave length of 370 nm. Using a narrow i-line filter (FWHM 12 nm at a center wavelength of 365 nm; 35 percent maximum transmission) provided significantly lower intensities (about one ninth of the intensity achieved with the broad filter). The demand for a wide filter can be explained by the strong line broadening due to the high operation pressure of the UHP mercury arc lamp.

\begin{figure*}[t]
    \includegraphics{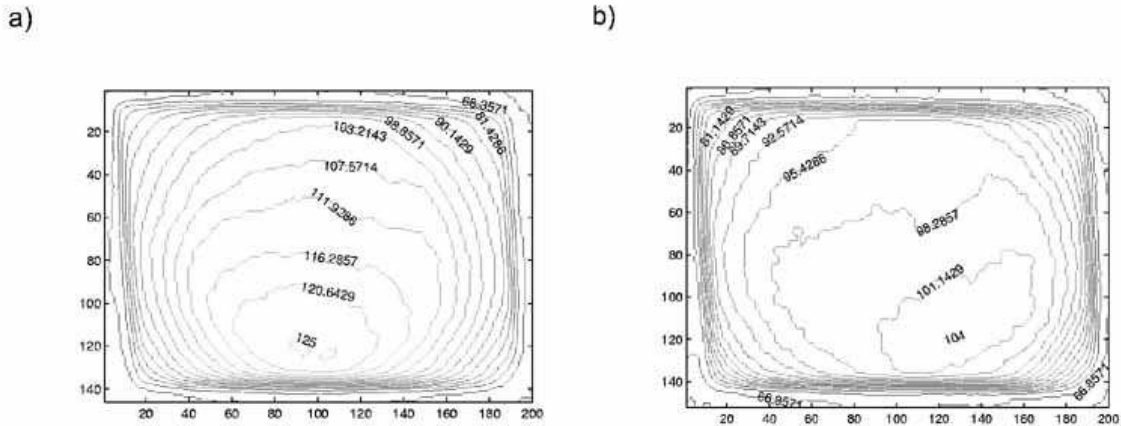}
  \caption{
Uniformity of illumination. (a) Intensity contour map before intensity leveling. (b) After intensity leveling. Using the tube lens, the  image of the DMD was projected onto a screen, without  the microscope objective in place,  and photographed with a digital camera. Vignetting from the microscope objective is neglected here  but this effect is small compared to vignetting of the tube lens.
  }
  \label{Labelname}
\end{figure*}

\subsection{Optical system performance testing with photochromic films}
\label{sec:OpticalPerformanceTesting}

Light-directed synthesis of DNA microarrays requires that the image is projected onto a substrate inside an inert reaction chamber, so that reactions can take place under a moisture free argon atmosphere. The synthesis substrate, a 0.17 mm thickness microscope cover glass, is forming the window of the reaction cell. Hence the image has to be projected onto the inner face of the window. For image focusing (see Sec.~\ref{Sec:ChromCorrection}) we use the small fraction of  green light which is reflected back from the imaging surface into the microscope. Applying a similar approach for contrast measurement is not practicable because the outer face of the cover glass contributes to back-reflection as well. Multiple reflections in the microscope system (e.g. from a beamsplitter) may degrade the image contrast further. 

The difference in total light intensity throughput of our optical system between a "full on" and a "full off" image exceeds 3000:1. We obtained this value using photochromic film for measurement. The aperture of the original projection objective built into the projector is much larger than the aperture of our setup, which is limited by the entrance diameter of the microscope objective. The tube lens is positioned at a large distance (about 165 mm) from the DMD-chip compared to the standard projector projection optics (about 25 mm).  Therefore in our setup the collection efficiency for diffusely scattered light is much smaller than in the original configuration, leading to a greatly reduced background at "full off". The unwanted background intensity increases significantly if patterns containing a large fraction of on-mirrors are displayed. This optical flare is  due to multiple reflections between optical surfaces. Use of optical elements with appropriate AR-coatings (in the 365 nm range) could possibly reduce this stray intensity. Photons leaking to unexposed features can produce base insertions and therefore affect the function of the DNA chip. The local contrast  between neighboring features is particularly important  for the light-directed DNA synthesis process.\cite{Kim2004} It is limited by the point spread function, the airy pattern of the microscope objective.  
 
The patterns used for microarray synthesis typically have an array structure with a pitch of 17 $\mu$m or less. To obtain an estimate of the stray light induced error rate we have measured the image contrast at high spatial frequencies.

We found  that the UV-sensitive films we already used for adjustment of the UV optics (see Sec.~\ref{Sec:ChromCorrection}) are very well suited for testing the performance of the photolithography system. For visual inspection of the patterns we used an optical microscope (Olympus IX81) equipped with an automated X-Y translational stage and with a high resolution CCD camera (C9100 EM-CCD, Hamamatsu Photonics). 

Patterns of regularly spaced line pairs (a pair thereby comprises a black and a white bar of equal width), were imaged onto photochromic film, prepared as described above. The spatial frequeny of the pattern was varied between 14 and 70 line pairs per millimeter (lp/mm). Using an exposure time scalebar (subsequently imaged nearby the line patterns) allowed us to determine the equivalent exposure time (due to stray light) within the dark lines. To obtain the relative stray light intensity this was set in relation to the exposure time (see Table~\ref{tab:StrayLightTab}).

\begin{table}[htbp]
	\centering
	\begin{ruledtabular}
		\begin{tabular}{ c c }
			  spatial frequency&relative stray light\\
			  (line pairs/mm)&intensity (percent)\\
			\hline
			70&10  \\
			35&5.5 \\
			28&3.2 \\
			21&2.2 \\
			14&0.5 \\
			
		\end{tabular}
		\end{ruledtabular}
	\caption{Relative stray light intensities in dependence of the spatial frequency of the line pairs. Stray light intensities were measured at the center of the unexposed lines. }
	\label{tab:StrayLightTab}
\end{table}

\begin{figure}[htp]
  
    \includegraphics[angle=0,width=7cm]{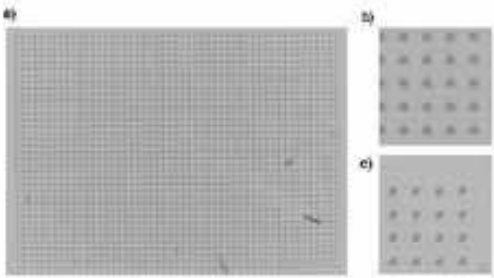}
  \caption{
Image distortion due to curvature of field. (a) A test pattern of 4$\times$4 pixel squares (pitch 20 pixels) covering the whole DMD area imaged onto photochromic film.  (b) Close-up view of the center region. The response of the photochromic material is asymptotically saturating in the center of the features. Therefore already a stray light intensity of 0.5 percent of the exposure intensity  results in a recognizable halo around the features. (c) Close-up of the upper right corner of the imaging field. The square features are radially distorted.   }
  \label{Labelname}
\end{figure}

The stray light intensity increases towards higher spatial frequencies, and is therefore setting an upper limit for the feature density in the light directed synthesis of microarrays. As described in Sec.~\ref{sec:LightDirectedSynthesis} a sufficiently high contrast to produce microarray arrays performing well in  mismatch discrimination has been achieved with a feature spacing of 17.5 $\mu$m. At larger feature densities the stray light from neighboring features will provoke base insertions, and hence will gradually decrease the quality of the synthesized DNA.

To demonstrate the effects of optical aberrations on the imaging performance, a pattern comprising of 4$\times$4 pixel squares with a pitch of 20 pixels was imaged onto photochromic material. A radial distortion of the square features is recognizable in Fig. 3(c) which was taken at a corner of the imaging field. As non-corrected curvature of field is supposed to be responsible for the distortion, we tried to improve image quality using a plan-corrected microscope objective. This, as well as using a narrower band pass filter to reduce chromatic aberrations didn't significantly improve the result. 
Another possible source for contrast impairment is the illumination system, which has been designed for a high light throughput. It may be possible to improve the optical aberrations, if this constraint is relaxed.

\begin{figure}[bp]
  
   \includegraphics[angle=0,width=8.5cm]{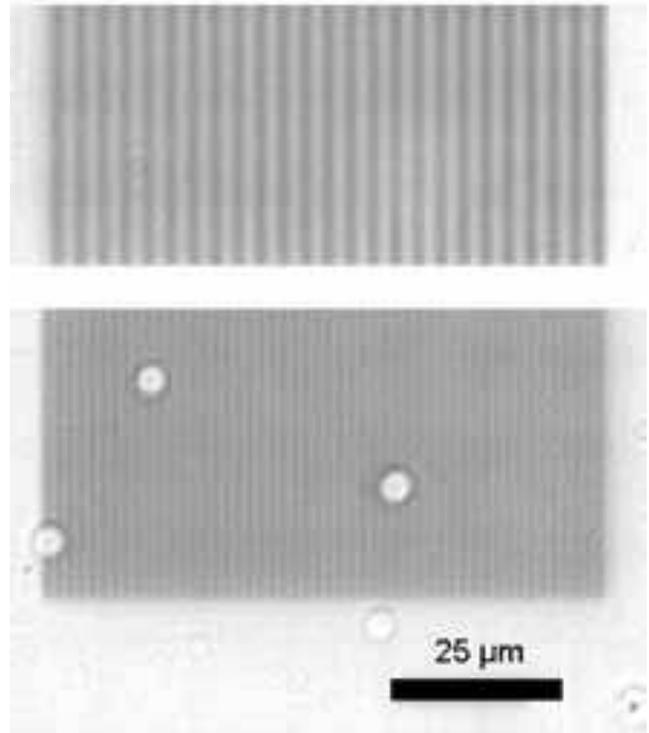}
  \caption{
Line patterns on photochromic film produced with a 20$\times$ (0.5 NA) Plan Neofluar objective. (a) Linewidth 1.7 $\mu$m (corresponding to a double row of micromirrors) (b) Linewidth 0.85 $\mu$m  (one single line of micromirrors).   
  }
  \label{Labelname}
\end{figure}

We found that using higher magnification objectives is possible. Using a  20$\times$ (0.5 NA) Plan NeoFluar (Zeiss), the total image size is reduced to 0.87$\times$0.65 mm\textsuperscript{2}. As shown in Fig. 4, spatial frequencies of  588 lp/mm (line width 0.85 $\mu$m) can clearly be resolved on the photochromic film. Due to reduced depth of focus, and non-corrected field curvature, this resolution can only be achieved in the center of the imaging area. At high spatial frequencies the measured contrast is additionally reduced due to the modulation transfer function of the inspection microscope optics. Therefore the contrast observed in Fig. 4 represents a lower limit. 

\begin{figure}[tp]
  
    \includegraphics[angle=0,width=8.5cm]{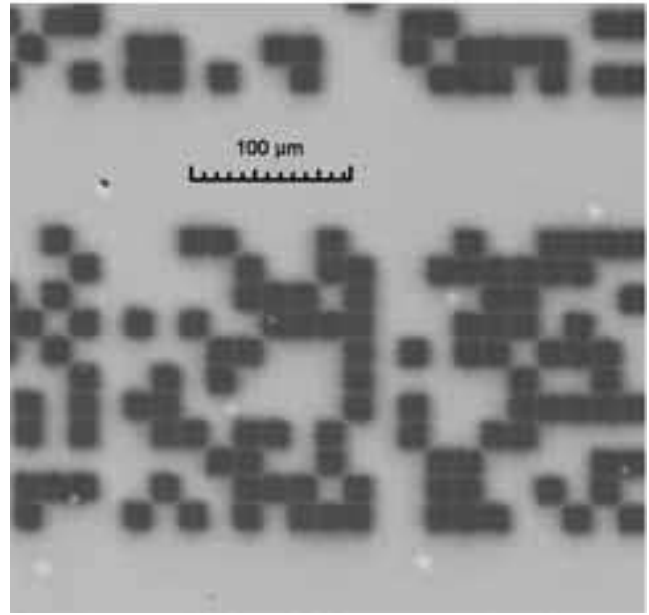}
  \caption{
A synthesis mask pattern, as used in light-directed synthesis of microarrays (compare with Fig. 6), is imaged onto photochromic film. The stray light affecting unexposed features depends on the density of exposed features nearby.  
  }
  \label{Labelname}
\end{figure}

To determine the stray light as it occurs during typical DNA microarray fabrication, a  synthesis mask pattern is projected onto the photochromic film (Fig. 5) with a fill factor (fraction of exposed pixels) of about 25 percent. We found that the stray light intensity reaching unexposed features is of the order of 0.5 percent of the exposure intensity. For unexposed features completely surrounded by exposed features we have measured a stray light intensity of 1.5 percent. These values are measured in the center of the unexposed features. At the edges of the features the stray light intensity can be significantly higher. 

The stray light induced error rate is even larger than stray light intensity suggests: During the long exposure that is necessary to drive the deprotection of the exposed features asymptotically towards saturation (necessary to prevent deletion errors), the number of stray light induced errors is growing almost linearly.

\subsection{Light-directed DNA microarray synthesis}
\label{sec:LightDirectedSynthesis}
DNA microarrays, also called DNA chips, are becoming increasingly important tools in genomics, molecular biology and medical diagnostics. A regular arrangement of microscopic spots (features) of  single stranded DNA molecules, which are bound to a solid support (for example to the surface of a microscope slide), is used to measure the expression levels of a large number of genes (i.e. the activity of these genes) simultaneously. The working principle of microarrays relies on the affinity of complementary strands of DNA, leading to formation of stable double helices. Fluorescently labeled target strands from the sample solution to be assayed bind to features containing complementary probe DNA sequences. By fluorescence detection methods these features and therefore the corresponding DNA sequences abundant in the sample solution can be identified.

Using MPLS high density oligonucleotide microarrays were manufactured in a light-directed approach similar as described by Singh-Gasson\cite{Singh-Gasson1999} and others.\cite{Nuwaysir2002,Kim2004,Luebke2002} Since this article is focused on the lithography system, we give here only a brief description of the experiments. A detailed description on design and performance of the DNA synthesis system will be published elsewhere.

In the  light-directed synthesis process\cite{Fodor1991}, UV photopatterning is used to control the coupling of chemically modified nucleotides (NPPOC-protected phosphoramidites\cite{Hasan97}) to the dangling ends of the growing substrate-bound oligonucleotide strands. Thousands of different DNA sequences can be synthesized simultaneously, arranged in an array of spatially separated domains - commonly called features. Photomasks are used to define the geometry of the microarray. The whole set of photomasks used in a synthesis also encodes the sequences of the DNA molecules to be synthesized in each one of the features. For example, the synthesis of a microarray containing 25mers (single stranded DNA molecules with 25 nucleotides) requires up to 4$\times$25=100 UV exposures - each one with a different photomask. (For simplicity we shall ignore here, that the masks are usually optimized, thus the synthesis requires a few less exposure steps.) The synthesis of a 25mer chip is subdivided into 25 nucleotide layers. For the completion of each layer all four nucleotide building blocks (A,C,G and T) have to be coupled consecutively, for example in the order A-C-G-T. The exposure step preceeding each coupling step, determines  at which locations (i.e. at which features) a given nucleotide building block is going to be coupled to the growing DNA strands in this layer. During exposure the photolabile NPPOC-protection group is cleaved from the terminal nucleotide of the DNA strand by UV irradiation. This photodeprotection reaction is exposing a previously covered (and therefore protected) hydroxyl group at the 5'-end, thus making the molecule available for coupling with the monomer building blocks added in the subsequent coupling step. Because the new terminal nucleotide is again featured with a photolabile protection group, no further couplings can occur to this DNA strand until the next UV exposure of the site.
The length of the probe molecules synthesized on the array is typically 15 to 30 nucleotides. Due to random errors, caused by stray light or incomplete coupling of the monomers, the yield Y of correctly synthesized n-mer oligonucleotides is restricted to $Y=E_s^{3n}\cdot E_c^n$. Here $E_s$ and $E_c$ denote the stepwise efficiencies for exposure steps respectively coupling steps. Depending on the different NPPOC phosphoramidites, the coupling efficiency $E_c$ has been reported to vary between 96 and 99 percent.\cite{Nuwaysir2002}  The stray light induced error rate, causing erroneous insertion of additional bases, is expressed in $E_s$, and depends on the geometry of the mask patterns (see Sec.~\ref{sec:OpticalPerformanceTesting}). With a stray light intensity of 0.5 percent  and the time course of photodeprotection (i.e. the fraction of deprotected DNA strands vs. duration of photodeprotection) the fraction of erroneously deprotected molecules per exposure step is estimated to be about one percent.
Therefore $E_s$ is estimated to have an average value of about 99 percent. Because stray light comes from the neighboring features, it will affect the synthesis of the feature in consideration, each time when it is not exposed - i.e. in total 3n times. With these efficiencies ($E_s=0.99$ and $E_c=0.97$) the yield Y of correctly synthesized sequences on a 25mer microarray is about 20 percent. For $E_s=0.95$ (corresponding to a stray light intensity of  2 percent) the yield would be significantly reduced to about 1 percent. Therefore a high local contrast ratio over the small distance separating neighboring features is crucial for successful light-directed fabrication of DNA microarrays. With a local image contrast ratio of the order of 1:100, the stray light-induced error rate is comparable to that of the limited coupling efficiency.

\begin{figure}[hbtp]
  
     \includegraphics[angle=0,width=8.5cm]{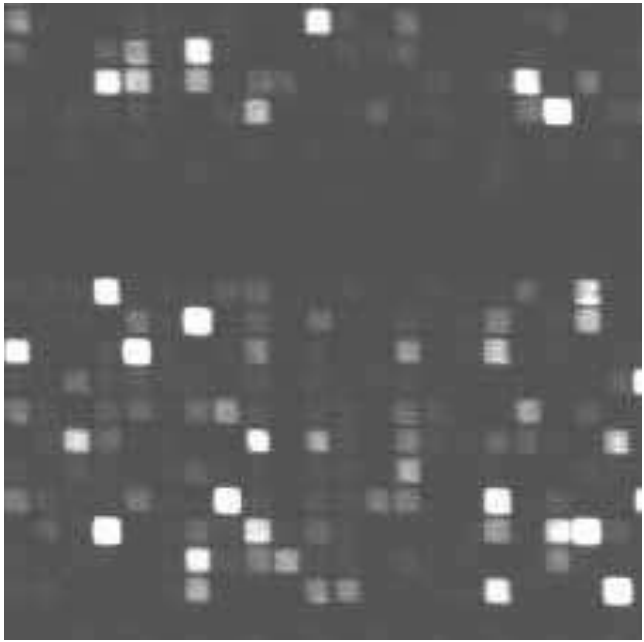}
  \caption{
Hybridized microarray (after washing off the unbound targets) as seen under the fluorescence microscope. Fluorescence intensity indicates the abundance of the target DNA corresponding to the particular feature. The feature size is 14 $\mu$m.  }
  \label{Labelname}
\end{figure}

\begin{figure*}[t]
     \includegraphics[angle=0,width=15cm]{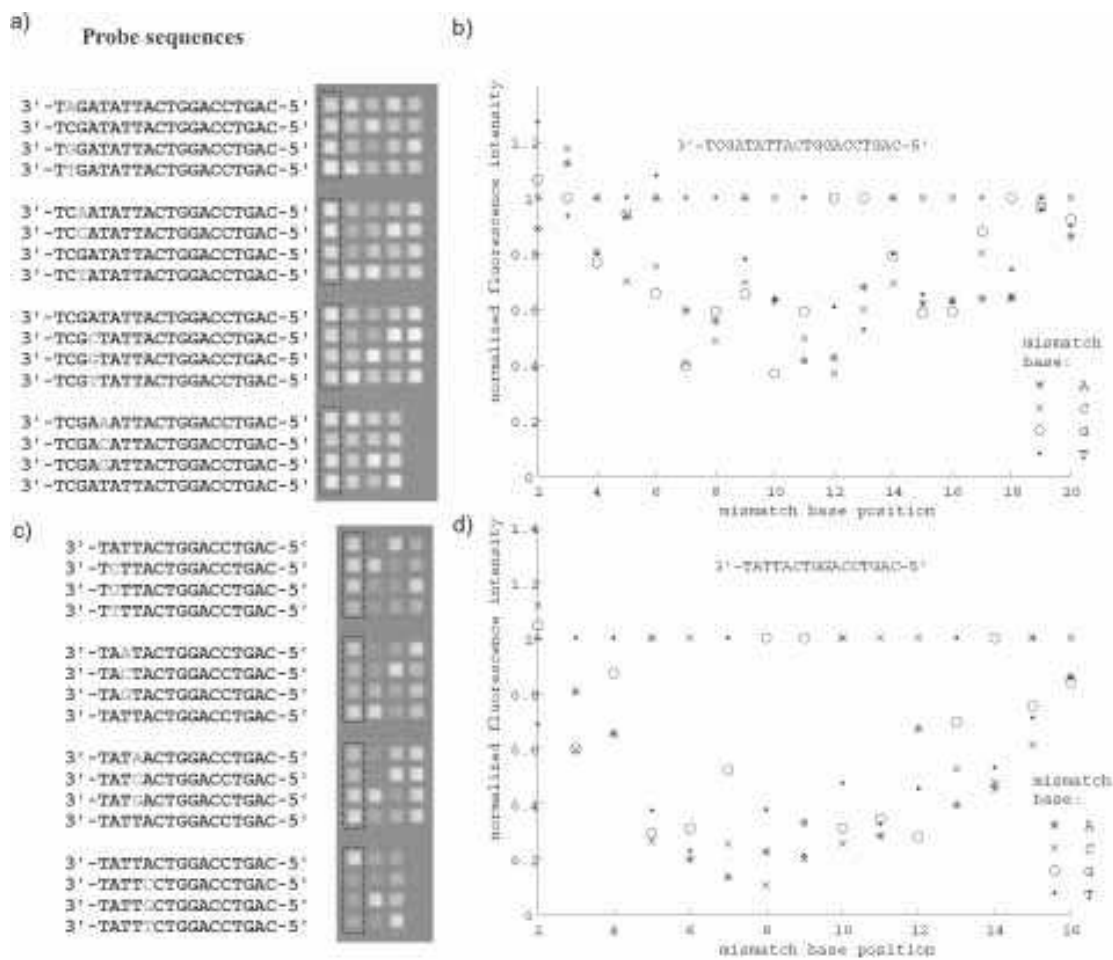}
  \caption{
Mismatch discrimination experiment.
We synthesized a microarray containing features with (almost) all single base mismatch variations of the 20mer sequence 3'-TCGATATTACTGGACCTGAC-5' (in the following called perfect match sequence - PM). In our experiment the fluorescently labeled (Cy3 at the 3'-end) target sequence  (which is completely complementary to PM) is hybridized onto the microarray. 
(a) The microarray features (size 14 $\mu$m, pitch 21 $\mu$m) shown in the fluorescence micrograph  are arranged in such a way that the position of the mismatch base is progressively moving from the 3'-end  to the 5'-end of the probe sequence (starting from the upper left of the feature set, running column by column to the lower right). 
At each base position of the probe sequence the base is varied (in the order A, C, T, G) resulting in a group of four features with one perfect match and three mismatches. For four of these groups (dashed frame) the corresponding probe sequences are provided. The mismatching bases (located at the base positions 2 to 5 in respect to the 3'-end of the probe sequence) are highlighted. The perfectly matching duplexes (PM features) provide the brightest signal and are clearly distinguishable from the mismatch features. 
In (b) the fluorescence intensity of the features (normalized with respect to the intensity of the PM features) is plotted vs. the mismatch position. It is noticeable that mismatch discrimination in the middle of the probe sequence (with a mismatch intensity of  0.4 - 0.8 times the perfect match intensity) is better than at the ends. 
(c) and (d) On the same chip an experiment with shortened probe sequences (four bases have been "removed" from the 3'-end of the 20mer sequence) has been performed. For the 16mer probe sequences the mismatch discrimination is significantly better (mismatch intensities are  0.1 to 0.5 times the PM-intensity) than for the 20mer probes.
  }
  \label{Labelname}
\end{figure*}
\begin{table*}[htbp]
	\centering
	\begin{ruledtabular}
		\begin{tabular}{ l l }
			Imaging optics&5$\times$0.25NA Fluar objective (Zeiss)  \\
			Exposure wavelength&370$\pm$17 nm \\
			Exposure intensity&76 mW/cm\textsuperscript{2} \\
			Pixel resolution&1024$\times$768 (XGA)\\
			Pixel size&3.5 $\mu$m \\
			Size of field&3.5 mm$\times$2.6 mm  \\
			Drift stability&$< $1 $\mu$m over 6 hours \\	
			Time required for synthesis of a 25mer chip & ca. 6.5 hours \\
			max. number of microarray features& 25000 (4$\times$4 pixel per feature and 1 pixel space) \\
			Useful area for DNA synthesis& ca. 80 percent of the DMD imaging field \\
			Reagent consumption for a 25mer synthesis & 30-40 mg of each NPPOC-phosphoramidite \\

		\end{tabular}
		\end{ruledtabular}
	\caption{Technical parameters of the MPLS DNA synthesis apparatus}
	\label{tab:TechParameters}
\end{table*}
Although most of the probe sequences synthesized contain synthesis errors (mostly deletions and insertions), we found that hybridization on our microarrays is highly specific (see Sec.~\ref{sec:HybAndMMDiscrimination}) and highly sensitive. 

Currently, with our microarray synthesizer, arrays comprising up to 25000 features  can be synthesized. With a feature size of 4$\times$4 pixel and a separation gap width of 1 pixel, 25 pixels are required per feature. A reduction of the feature size to 3$\times$3 or 2$\times$2 pixels (taking into account a separating gap of 1 pixel between the features) would allow 49000 respectively 87000 features. Preliminary results have shown that a reduction to 3$\times$3 pixels is possible, but due to increased stray light  the quality of the synthesized DNA is expected to decrease. This is in accordance with results from Kim \textit{et al.}.\cite{Kim03} In a numerical simulation of the synthesis process they found that the purity of the oligomers is greatly improved (significantly less insertions) just by increasing the spacing between features. The effects of stray light on the quality of the DNA and on the function of the microarray are currently under investigation.

The synthesis of a 25mer chip requires roughly 75 optimized exposure steps and subsequent base additions. With an exposure time of 90 s per deprotection step the total exposure time amounts to  roughly 2 hours - about one third of the total synthesis time. The volume of the reaction cell is 50 $\mu$l, the cell is formed by a streamlined cut-out in a sheet of PDMS silicone rubber.  An amount of 60 $\mu$l of NPPOC-amidite coupling solution (prior to coupling mixed with an equal amount of  tetrazole activator) is used per synthesis step. 
To prevent unduly losses of the expensive reagents in the tubing resp. in the valve system, pure acetonitrile is used to press the amidite solution into the reaction cell. In a 25mer synthesis the consumption of amidite is only 30-40 mg of each of the four amidites. To enhance the coupling reaction agitation is applied by a series of pulses. The coupling solution is transported by a few mm at each pulse.

The technical parameters of the MPLS DNA synthesis apparatus are summarized in Table~\ref{tab:TechParameters}.

\subsection{Microarray hybridization and mismatch discrimination}
\label{sec:HybAndMMDiscrimination}
For the hybridization assay a salt buffer solution containing the nucleic acid sample to be assayed, is applied onto the microarray surface. The target DNA molecules, priorly labeled with the fluorescent dye Cy3, can freely diffuse and interact with the surface bound probe DNA molecules (previously synthesized on the microarray).  Stable binding (hybridization)  between probe and target strands  occurs if strands are complementary to each other over a sufficiently long stretch. Owing to the slow diffusion process, hybridization is allowed to take place over several hours. Subsequently, washing removes unbound or weakly bound molecules from the surface. After washing, remaining target molecules are those bound to features containing complementary probes sequences. The features of the microarray therefore show a more or less bright fluorescence depending on the abundance of the complementary target molecules in the sample. Identification of a feature is based on its position in the regular grid of the microarray. Fluorescence microscopy (Olympus IX-81 equipped with a M\"arzh\"auser X-Y translational stage and  Hamamatsu EM-CCD camera) is used for readout of the microarray data. In Fig. 6 a small section of a hybridized microarray is shown. Despite the relatively large stray light intensities at the edge of the features, the fluorescence intensity of the features is almost constant throughout the feature area and the 3.5 $\mu$m gaps separating the features are clearly visible.

We found that our microarrays perform well in mismatch discrimination. We synthesized a chip containing a set of features comprising (almost) all single base mismatches of a 20mer (respectively of a 16mer) probe sequence (see Fig. 7). At each base position, proceeding from the surface-bound 3'-end  to the 5'-end of the probe sequence, the nucleobase is varied in the order A, C, G and T, thus  providing a set of 4 features (with one perfect match and three mismatches) for each base position. Hybridization with the Cy3-labeled target oligonucleotide (concentration 1 nM, at 37$^{\circ}$C in a hybridization buffer according to Nuwaysir \textit{et al.}\cite{Nuwaysir2002}) demonstrates the microarrays ability to discriminate the programmed single base mismatches from perfect matching sequences. Mismatches in the middle of the probe sequence affect hybridization more than mismatches located near the ends of the sequence (Figs. 7(b) and 7(d)). Reduction of the probe length to 16 bases reduces the duplex stability and significantly improves the mismatch discrimination ability (see Figs. 7(c) and 7(d)).

\begin{figure}[htp]
  
     \includegraphics[angle=0,width=8.5cm]{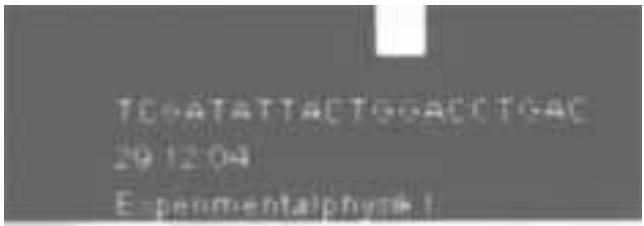}
  \caption{
The text shown here is written in DNA. Under fluorescence exitation the labeled target DNA strand  hybridized to the surface bound probe strand (forming a stable double helix) lights up brightly. As the target DNA only binds to its complementary probe strand this shows that over the synthesis time of 6 hours the setup must have been mechanically stable within considerably less than 3.5 $\mu$m (which is the linewidth of the text). }
  \label{Labelname}
\end{figure}
The high mechanical stability of the photolithography system is demonstrated in Fig. 8. The successful hybridization at "text features" with a line width of 3.5 $\mu$m  indicates that during the synthesis the alignment of the patterns has been maintained within about one micron. 
\begin{figure*}[ht]
      \includegraphics[angle=0,width=17cm]{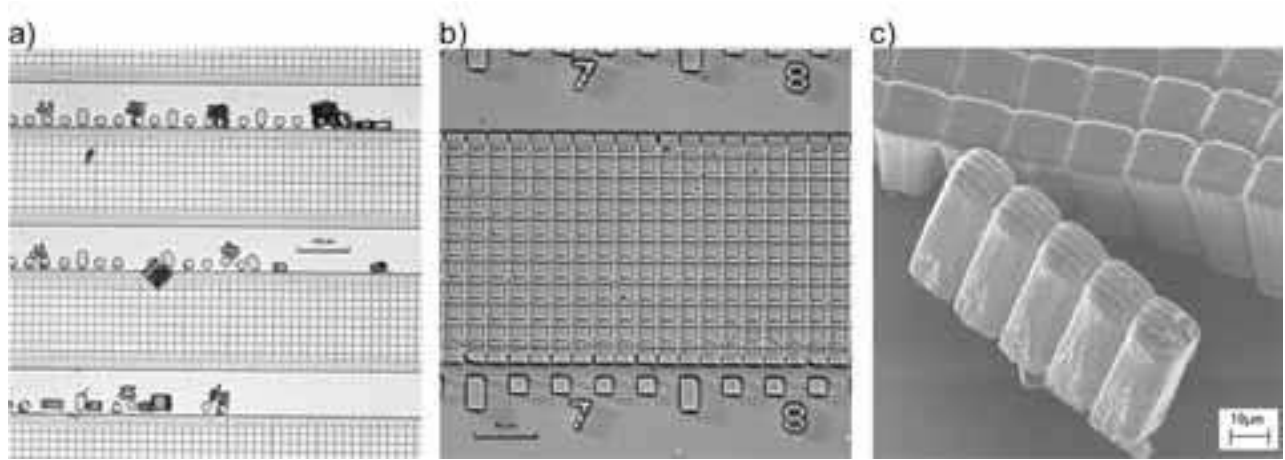}
  \caption{
  MPLS-generated pattern in  SU-8 photoresist - film thickness: 40 $\mu$m. 
(a)   The pattern geometry is the same as in Fig. 6. Each block corresponds to 4$\times$4 Pixels, the separation gap is one pixel wide.  Some of the letters are lying sideways on the surface, demonstrating that an aspect ratio of 1:10 is achievable with MPLS (scalebar 100 $\mu$m).
(b) Micrograph of the same pattern as in (a). Even small features like the number "seven" with a line width of only 3.5 $\mu$m (corresponding to a single micromirror, the pixelation is clearly visible) are reproduced in the photoresist (scalebar 50 $\mu$m). 
(c) Electron micrograph of the photoresist structures. Due to relatively poor surface adhesion of the photoresist the structures have partly detached from the glass surface (scalebar 10 $\mu$m).
  }
  \label{Labelname}
\end{figure*}

\subsection{Photoresist patterning}

To demonstrate the application of MPLS as a highly flexible photolithography system we have produced microstructures in SU-8 photoresist (Microchem Corp.). The SU-8 50 negative resist was spincoated on microscope slides at 3000 rpm, resulting in a thickness of 40 $\mu$m.
For the experiment a pattern of tightly spaced square-features 14 $\mu$m in size, separated by 3.5 $\mu$m gaps was used (this is actually the same pattern format as used for the microarray synthesis - compare to Fig. 6). Processing was accomplished according to the processing guidelines of the manufacturer.
The photoresist, which is sensitive between 350 and 400 nm has been exposed to UV light for different times, ranging from 5 s to 25 s. After postbaking and developing the resulting microstructures were imaged using  a research microscope as described above. We found 15 s to be an appropriate exposure time. Fig. 9 illustrates the high quality of the resulting microstructures. As one can see in Fig. 9(b), features as small as 3.5 $\mu$m (the line width of the number "7") have been reproduced very well. The aspect ratio of the structures as can be seen in Fig. 9(c), is about 1:10.


\section{Conclusions}
We have demonstrated and characterized a versatile microscope projection photolithography system, which is simple to set up. The apparatus, built from a video-projector and a standard commercial inverted microscope, has been employed successfully for the light-directed synthesis of DNA microarrays of reduced size. Size reduction can be expected to increase the sensitivity of the array.\cite{Wei2005} The fabrication process, during which a multitude of up to 100 masks is projected, typically requires 6 to 8 hours. Within this time the setup is mechanically stable, no drift could be detected during a large number of synthesis experiments. The smallest achievable line-width for DNA is 3.5 $\mu$m with a 5$\times$ objective. The quality and applicability of the microarrays synthesized have been demonstrated in an experiment designed to study single base mismatch discrimination ability. We have also shown the capability of the setup for photoresist patterning. Here we have achieved a minimum feature size of 3.5 $\mu$m with a surprisingly high aspect ratio of about 1:10. On UV-sensitive photochromic films, we achieved submicron resolution using a 20$\times$ microscope objective, albeit with a reduced contrast. The setup further has the capability for grey scale patterns and for dynamic mask projection, which opens up a large range of new applications. An interesting possibility in the field of DNA microarrays is an automated, recursive optimization of the DNA sequences employed. As the array can be tailored to the specific sample, such a procedure should greatly enhance the confidence of an array measurement. Another example for the future potential of DMD based microprojection is the application in micro-stereolithography for the fabrication of complex three-dimensional microstructures.\cite{Sun2005} 

\newpage
\section{References}



\end{document}